\documentclass[prl,aps,twocolumn,floatfix,superscriptaddress]{revtex4-1}

\usepackage{amstext}
\usepackage{amssymb,amsmath} 
\usepackage{bbold}
\usepackage{hyperref}
\usepackage{graphicx} 
\usepackage{multirow} 

\begin{document}
\title{More randomness from noisy sources}

\date{\today}

\author{Jean-Daniel Bancal}
\affiliation{Centre for Quantum Technologies, National University of Singapore, 3 Science drive 2, Singapore 117543}
\author{Valerio Scarani}
\affiliation{Centre for Quantum Technologies, National University of Singapore, 3 Science drive 2, Singapore 117543}
\affiliation{Department of Physics, National University of Singapore, 2 Science drive 3, Singapore 117542}

\begin{abstract}
Bell experiments can be used to generate private random numbers. An ideal Bell experiment would involve measuring a state of two maximally entangled qubits, but in practice any state produced is subject to noise. Here we consider how the techniques presented in~\cite{Bancal14} and~\cite{Olmo14}, i.e. using an optimized Bell inequality, and taking advantage of the fact that the device provider is not our adversary, can be used to improve the rate of randomness generation in Bell-like tests performed on singlet states subject to either white or dephasing noise.
\end{abstract}

\maketitle

\section{Introduction}\noindent
It is well known that the violation of a Bell inequality rules out the possibility for the outcomes of a Bell-type experiment to be known in advance~\cite{review}. Therefore, these outcomes are certifiably unpredictable. Recent works have shown that the uncertainty present in these outcomes can be quantified, thus allowing one to lower bound the number of random bits that can be extracted from a given Bell-type experiment~\cite{Colbeck,Pironio10}.

This possibility has given rise to a variety of randomness-related studies based on a similarly varied set of working assumptions. For instance, many works considered the case in which the adversary (the actor for whom outcomes are to be certifiably unpredictable) is allowed to distribute the quantum state measured by the authorized parties, and keep a purification of this state. Under this assumption, it was shown that \emph{randomness expansion} is possible: if the user holds a secret string of finite length, he can expand it into a longer one~\cite{VV11}, or, in principle, even an infinite one~\cite{coudron,millershi}.

Also, the outcomes observed by the user can be certified to contain some amount of randomness even when the adversary, in addition to distributing the state, holds partial information about the initial random string of the user~\cite{CR}. This possibility, refered to as \emph{randomness amplification}, was proved recently for initial randomness issued from a generic min-entropy source~\cite{bouda,shi} after a series of partial results~\cite{Gallego,thinh}.

These results show the full power of quantum certification in principle. However, when it comes to realizing such protocols, a number of questions arise. For instance, in which practical situation would one wish to expand a random string if we already have access to a source that can produce initial random strings? Also, in the context of randomness amplification, it is unclear in which meaningful situation the dependence would exist at all but be bounded. If the adversary is allowed to tamper with the devices, for instance, or even to produce them, then he may have hidden some kind of emitter inside the boxes, in order to retrieve all numbers produced by the boxes (which otherwise work as expected). This simple possibility would compromise any certification of randomness.

For these reasons, in the design and assessment of practical realization of randomness protocols
, it is very reasonable to work under the assumption that the adversary has no access to the devices used by the authorized partners. This \emph{trusted provider} assumption was already introduced in the context of randomness protocols in~\cite{Pironio11}, where it was shown that it restricts the adversary to hold only classical side information (i.e. he cannot hold a purification of the quantum state). Note that this contrasts with the case of quantum key distribution (QKD): practical QKD also requires the trusted provider assumption, for the same reason as mentione above, however the adversary can still hold a purification in this case since the quantum state passes in his hands. Another consequence mentioned in~\cite{Pironio11} is that the initial string used by the user to choose settings for his Bell test need not be private, but can be fully known in advance by the adversary. One thus speaks of \emph{randomness generation} in this context.

It was shown in~\cite{Bancal14} that additional randomness can be certified under the trusted provider assumption compared to that granted by randomness expansion protocols, by extracting randomness from all the settings. Moreover, this same paper as well as~\cite{Olmo14} demonstrated that Bell-like inequalities that certify more randomness than usual Bell inequalities (like e.g. CHSH) can be derived from knowledge of the full correlations. In this paper, we analyse the advantage provided by these techniques when the quantum state measured by the user is a singlet states mixed either with white or dephasing noise. White noise typically describes the effect of many small errors in a setup whereas dephasing noise is the dominant noise in SPDC-based sources when the pump power is low. The case of white noise was already partially studied in both~\cite{Bancal14} and~\cite{Olmo14}. The analysis given here gathers the information presented in both studies and provides a comparison with the dephasing noise case.

Even though our analysis relies on the trusted provider assumption, it is worth noting that some of the results obtained here could also apply to more general adversaries; we refer to~\cite{Law14} for a concise review of adversarial classes relevant to randomness protocols.

For the present paper, we assume that the source emits exactly one pair of particles per unit time and that these are detected with certainty. The case of finite detection efficiency was studied in~\cite{Bancal14}, in absence of noise; when the emission is not heralded, more effects come into play, see e.g.~\cite{vivoli}.

Another assumption that we make here is that the devices used by the user are i.i.d. and that he can use as many of them as he wants. We thus focus on the rate of randomness generation, defined as the number of random bits generated in each use of the devices.

\section{Randomness analysis}\noindent
We consider here a usual Bell-type experiment performed by a user~\cite{review}. At each round, the user chooses some inputs $x,y$ for his two devices to use as measurement settings, and observes their outcomes $a,b$. The i.i.d. behavior of the boxes follows the quantum conditional probability $P(a,b|x,y)\in\mathcal{Q}$.

In general, these correlation can admit a decomposition $\{q_\lambda, P_\lambda\}$ such that
\begin{equation}\label{eq:decomp}
P(ab|xy)=\sum_\lambda q_\lambda P_\lambda(ab|xy)
\end{equation}
with $q_\lambda\geq0$, $\sum_\lambda q_\lambda = 1$, $P_\lambda(ab|xy)\in\mathcal{Q}$. When this decomposition is not trivial, by knowing in each round which value of $\lambda$ corresponds to the realization of the box, the adversary can hold a more precise decription of the box's behavior for that run, as given by $P_\lambda(ab|xy)$.

Following~\cite{Bancal14, Olmo14}, we thus define the adversary's guessing probability on the outcomes observed by the user when using settings $x,y$ and in presence of the decomposition $\{q_\lambda,P_\lambda\}$ as
\begin{equation}
G_{x,y}(\{q_\lambda,P_\lambda\}) = \sum_\lambda q_\lambda \max_{a,b} P_\lambda(ab|xy).
\end{equation}
The average guessing probability when settings are chosen with probability $p(x,y)$ is then the maximum of
\begin{equation}
G(P) = \sum_{xy} p(x,y) \sum_\lambda q_\lambda \max_{a,b} P_\lambda(ab|xy)
\end{equation}
over all decompositions~\eqref{eq:decomp} compatible with the correlations $P(ab|xy)$.

It was shown in~\cite{Bancal14} that this quantity can be upper bounded by considering an SDP (Semidefinite Program) relaxation of the set of quantum correlations~\cite{NPA}. In the following section, we thus use this program to evaluate the rate, as given by the min entropy
\begin{equation}
H_\text{min}(P) = -\log_2(G(P)),
\end{equation}
at which random bits are generated in the experiment.

Note that the particular case of this optimization where randomness is extracted from a fixed choice of settings ($p_{xy}=\delta_{x,x_0}\delta_{y,y_0}$), or where the outcomes of different settings are allowed to by guessed with different decomposition, was also presented independently in~\cite{Olmo14}.

In the following we compare three quantities:
\begin{enumerate}\label{itemize:cases}
\item The rate of randomness obtained from a fixed set of settings as certified by a CHSH violation.
\item The rate of randomness obtained from a fixed set of settings as certified by an optimized Bell-type expression.
\item The rate of randomness obtained when using all settings with the same probability as certified by an optimized Bell-type expression.
\end{enumerate}

Note that here we consider extracting randomness from the pair of outcomes $(a,b)$ rather than from the outcome of a single party. A similar computation could be done by taking only one party's outcome into consideration, but would result in a lower rate. Also, for the first two quantities, the fixed set of settings is chosen as to maximize the rate of randomness.

For all results presented next, the numerical computations were performed using the relaxation of the SDP hierarchy at local level 2~\cite{Moroder}.

\subsection{White noise}\noindent
First, let us consider the case in which the measured state is
\begin{equation}
\rho(V) = V|\phi^+\rangle\langle\phi^+|+(1-V)\openone/4,
\end{equation}
for some visibility $V$. The settings which provide the largest violation $2\sqrt{2}V$ of the CHSH inequality
\begin{equation}
S=\langle A_0B_0 \rangle + \langle A_0B_1 \rangle + \langle A_1B_0 \rangle - \langle A_1B_1 \rangle \leq 2,
\end{equation}
where $A_x$, $B_y$ are Alice's and Bob's observables, are the same for all $V$:
\begin{equation}
A_0 = \sigma_z,\ \  A_1 = \sigma_x,\ \  B_y = \frac{\sigma_z+(-1)^y \sigma_x}2.
\end{equation}

\begin{figure}
\includegraphics[width=0.5\textwidth]{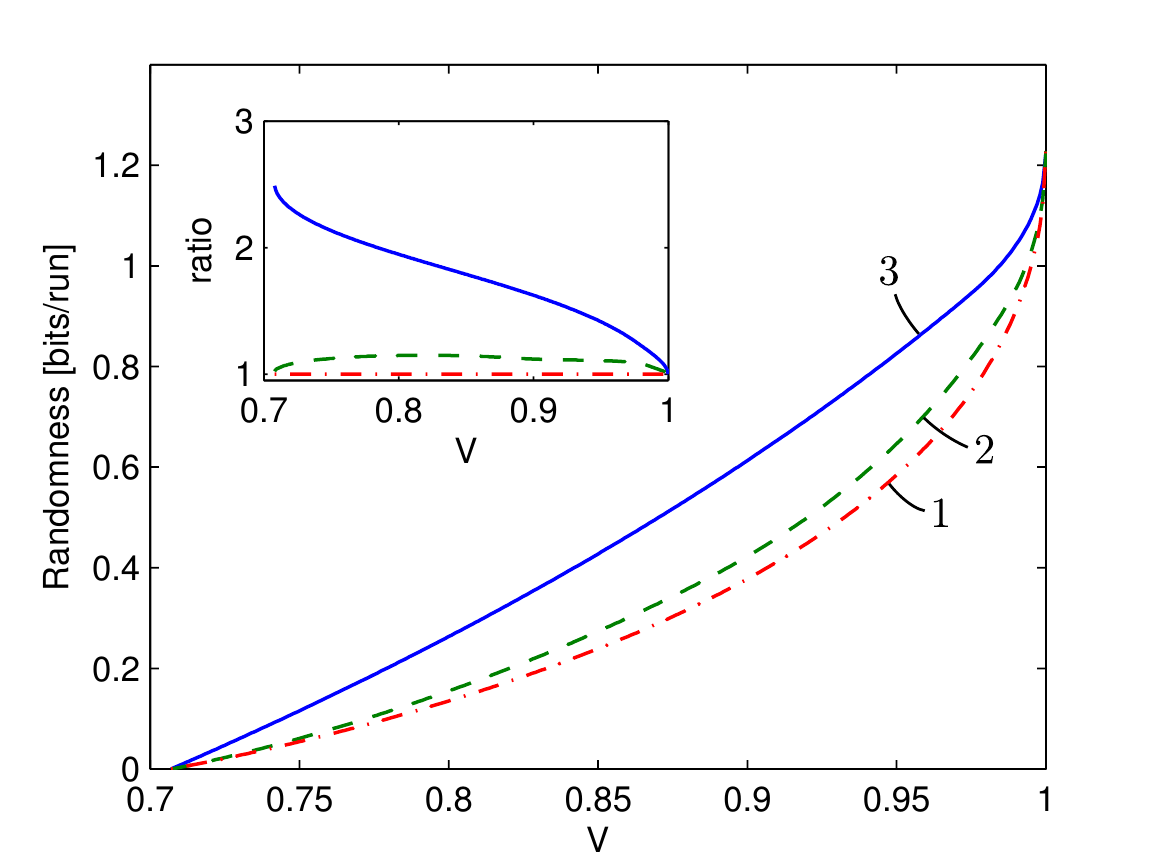}
\caption{Rate of private randomness generation certified by the measure 1, 2 and 3 for a singlet state mixed with white noise. The inset presents the ratio of the curves to the lowest one.}
\label{fig:figV}
\end{figure}

We thus computed for this state and settings the three different rates of randomness mentioned above. The result is presented in Figure~\ref{fig:figV}.

The randomness rate obtained in case 2 (middle curve) can be certified with the help of the following Bell expression:
\begin{equation}\label{eq:newineq}
\alpha \langle A_0B_0 \rangle + \langle A_0B_1 \rangle + \langle A_1B_0 \rangle - \beta \langle A_1B_1 \rangle,
\end{equation}
where the values of $\alpha$ and $\beta$ depend on $V$ (see~\cite{Olmo14} for a description of this dependence). 
The inset in Figure~\ref{fig:figV} shows that the advantage provided by using this optimized Bell expression is however quite limited.

The largest amount of randomness is obtained in case 3 when considering the outcomes observed when all settings are used with the same probability (i.e. $p(x,y)=1/4$). As mentioned in~\cite{Bancal14}, the improvement, of the order of a factor of 2, is certified with the usual CHSH inequality.

\subsection{Dephasing noise}\noindent
Second, we consider measurement of the state
\begin{equation}
\rho(p) = p|\phi^+\rangle\langle\phi^+|+(1-p)(|00\rangle\langle00|+|11\rangle\langle11|).
\end{equation}
The optimal violation of the CHSH inequality by this state, $S=2\sqrt{1+p^2}$, is provided by using the following settings~\cite{Horodecki}:
\begin{equation}
\begin{split}
A_0 &= \sigma_z,\ \  A_1 = \sigma_x,\ \\\
B_y &= \cos\chi\,\sigma_z + (-1)^y \sin\chi\,\sigma_x,
\end{split}
\end{equation}
with $\chi=\arctan(p)$.

\begin{figure}
\includegraphics[width=0.5\textwidth]{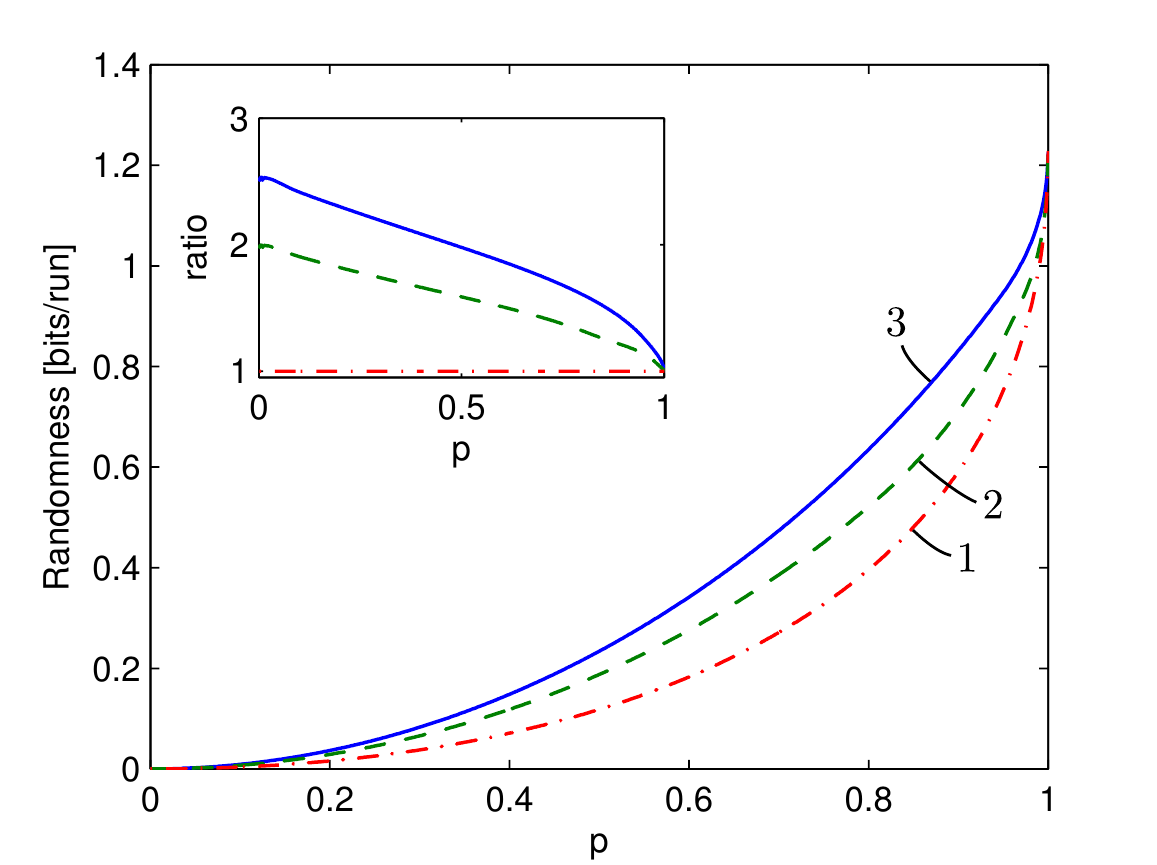}
\caption{Rate of private randomness generation certified by the measure 1, 2 and 3 for a singlet state mixed with dephasing noise. The inset presents the ratio of the curves to the lowest one.}
\label{fig:figp}
\end{figure}

The three randomness rates obtained with these state and settings are presented in Figure~\ref{fig:figp}. Similarly to the previous case, strictly more randomness can be certified in case 3 than in case 2, and in case 2 than in case 1. The inequality that certifies the largest amount of randomness is again CHSH in case 3, and a different inequality in case 2. One can check that this inequality, however, beyond being a correlation inequality presents no special symmetry. In particular, it is not of the form~\eqref{eq:newineq}. Nevertheless, we note that when the randomness is extracted from a single set of settings, using an optimized inequality provides a larger advantage for this dephasing noise than it did in the case of mixture with white noise (as shown in the inset of Figure~\ref{fig:figp}).

\section{Conclusion}\noindent
We have presented an application of the techniques presented in~\cite{Bancal14,Olmo14} to the case where the measured state is a singlet mixed with either white noise or dephasing noise. While a significant advantage in terms of randomness rate can be obtained in both cases when randomness is extracted uniformly from all settings, the advantage for extraction from a fixed choice of settings is much more significant in the case of dephasing noise.

In a practical experiment, characteristics of both white and dephasing noise are expected to appear~\cite{Kofler13}, as well as various other kind of noises and imperfections \cite{vivoli}. The present analysis is not meant to exhaust all the parameter space of a realistic experiment; but it should be clear that the techniques used here can be extended to describe experiments with all their features.

We have focused here on the asymptotic rate of randomness generation. It would be interesting to extend our analysis to take into account finite statistics, maybe in a way similar to~\cite{Pironio11} or~\cite{Zhang11}. This would allow one to quantify how many random bits can be extracted from a Bell experiment which involves only a finite number of rounds.

\section{Acknowledgements}
This work was supported by the National Research
Foundation (partly through the Academic Research Fund
Tier 3 MOE2012-T3-1-009) and the Ministry of Education, Singapore.

\end{document}